\documentclass[]{spie}  %

\usepackage{aas_macros}
\usepackage{amsmath,amsfonts,amssymb}
\usepackage{color}
\usepackage{caption}
\usepackage{subcaption}
\usepackage{graphicx}
\usepackage[colorlinks=true, allcolors=blue]{hyperref}
\usepackage{xspace}

\title{Black Hole Explorer Mission Development in Japan}

\newcommand{\heriotwatt}{Institute of Sensors, Signals and Systems, Heriot-Watt University, Edinburgh, EH14 4AS, United Kingdom}
\newcommand{\naojmizusawa}{Mizusawa VLBI Observatory, National Astronomical Observatory of Japan, Iwate 023-0861, Japan}
\newcommand{\mithaystack}{Massachusetts Institute of Technology Haystack Observatory, Westford, MA 01886, USA}
\newcommand{\cfa}{Center for Astrophysics $|$ Harvard \& Smithsonian, Cambridge, MA 02138, USA}
\newcommand{\harvardbhi}{Black Hole Initiative at Harvard University, Cambridge, MA 02138, USA}
\newcommand{\yamaguchi}{Graduate School of Sciences and Technology for Innovation, Yamaguchi University, Yamaguchi 753-8512, Japan}
\newcommand{\isasjaxa}{Institute of Space and Astronautical Science, Japan Aerospace Exploration Agency, Kanagawa 252-5210, Japan}
\newcommand{\sokendaijaxa}{Department of Space and Astronautical Science, The Graduate University for Advanced Studies, SOKENDAI, Kanagawa, 229-8510, Japan}
\newcommand{\harvardhistory}{Department of History of Science, Harvard University, Cambridge, MA 02138, USA}
\newcommand{\harvardphysics}{Department of Physics, Harvard University, Cambridge, MA 02138, USA}
\newcommand{\toyo}{Natural Science Laboratory, Toyo University, Tokyo 112-8606, Japan}

\newcommand{\utokyoastro}{Department of Astronomy, Graduate School of Science, The University of Tokyo, Tokyo 113-0033, Japan}
\newcommand{\niigata}{Graduate School of Science and Technology, Niigata University, Niigata 950-2181, Japan}
\newcommand{\naojatc}{Advanced Technology Center, National Astronomical Observatory of Japan, Tokyo 181-8588, Japan}
\newcommand{\utsukuba}{Center for Computational Sciences, University of Tsukuba, Ibaraki 305-8577, Japan}
\newcommand{\gifu}{Faculty of Engineering, Gifu University, Gifu 501-1193, Japan}

\newcommand{\ncu}{The Graduate School of Science, Nagoya City University, Aichi 467-0001, Japan}

\newcommand{\musashino}{Department of Mathematical Engineering, Faculty of Engineering, Musashino University, Tokyo 135-8181, Japan}
\newcommand{\nitichinoseki}{National Institute of Technology, Ichinoseki College, Iwate 021-8511, Japan}
\newcommand{\arizona}{Steward Observatory and Department of Astronomy, University of Arizona, Tucson, AZ 85721, USA}
\newcommand{\shao}{Shanghai Astronomical Observatory, Chinese Academy of Sciences, Shanghai 200030, P. R. China}
\newcommand{\bunkyo}{Faculty of Education, Bunkyo University, Saitama 343-8511, Japan}

\author[1,2,3,4]{Kazunori Akiyama}
    \affil[1]{\heriotwatt}
    \affil[2]{\naojmizusawa}
    \affil[3]{\mithaystack}
    \affil[4]{\cfa}
\author[2,5]{Mareki Honma}
    \affil[5]{\utokyoastro}
\author[6,7]{Akihiro Doi}
    \affil[6]{\isasjaxa}
    \affil[7]{\sokendaijaxa}
\author[8]{Shoko Koyama}
    \affil[8]{\niigata}
\author[9]{Kotaro Niinuma}
    \affil[9]{\yamaguchi}
\author[10,2]{Kazuhiro Hada}
    \affil[10]{\ncu}
\author[9]{Yuto Akiyama}
\author[4]{Rebecca Baturin}
\author[11,12,13]{Peter Galison}
    \affil[11]{\harvardbhi}
    \affil[12]{\harvardphysics}
    \affil[13]{\harvardhistory}
\author[4]{Paul Grimes}
\author[14]{Yoshiaki Hagiwara}
    \affil[14]{\toyo}
\author[15,2]{Takayuki J. Hayashi}
    \affil[15]{\bunkyo}
\author[16]{Aya E. Higuchi}
    \affil[16]{\musashino}
\author[4]{Janice Houston}
\author[4,11]{Michael D. Johnson}
\author[17]{Tomohisa Kawashima}
    \affil[17]{\nitichinoseki}
\author[18]{Daniel P. Marrone}
    \affil[18]{\arizona}
\author[19]{Yosuke Murayama}
    \affil[19]{\naojatc}
\author[20]{Ken Ohsuga}
    \affil[20]{\utsukuba}
\author[4,11]{Hannah Rana}
\author[21]{Hidetoshi Sano}
    \affil[21]{\gifu}
\author[4]{Edward Tong}
\author[22,20]{Yuh Tsunetoe}
    \affil[22]{\shao}
\author[19]{Yoshinori Uzawa}

\authorinfo{Further author information: (Send correspondence to K.A.)\\K.A.: E-mail: k.akiyama@hw.ac.uk}

\def\M87{M87$^*$\xspace}
\def\m87{M87$^*$\xspace}

\begin{document} 
\maketitle

\begin{abstract}
The Black Hole Explorer (BHEX) is a next-generation space very-long-baseline interferometry (VLBI) mission concept that will extend existing ground-based millimeter/submillimeter VLBI arrays to space. The Japanese astronomical community has contributed to BHEX mission development through the BHEX Japan Consortium, established in 2023.
This paper provides a high-level summary of progress in Japan since 2024, including the establishment of the Black Hole Explorer Working Group (BHEX WG) at the Institute of Space and Astronautical Science (ISAS), JAXA, to conduct the Japanese side of the Pre-Phase~A mission studies.
We outline recent advances in key instrument technologies, including concept design studies of a 4.5\,K closed-cycle mechanical cryocooler and prototype development of an ultra-wideband 300\,GHz Superconductor--Insulator--Superconductor (SIS) mixer for BHEX.
We also describe ongoing upgrades to Japan's ground infrastructure to support 86\,GHz observations with VERA and simultaneous 86+230\,GHz observations with the Nobeyama 45\,m Telescope.
\end{abstract}


\section{Introduction}
\label{sec:intro}  %
The Black Hole Explorer (BHEX)\cite{BHEX_Johnson_2024}\footnote{\url{https://www.blackholeexplorer.org/}, accessed on June 30, 2026} is a next-generation space very-long-baseline interferometry (VLBI) mission concept currently under development within NASA's Astrophysics Small Explorer (SMEX) program.
The primary goal of BHEX is to achieve the first direct detection of a black hole's ``photon ring,'' which is composed of light rays that have orbited the black hole before escaping. Because the photon ring traces a narrow region of space immediately outside the black hole's event horizon, it serves as a unique probe of spacetime, enabling direct measurements of the black hole's spin and a precise probe of strong-field gravity using light\cite{Johnson_2020, Gralla_2020, Palumbo_2023, Lupsasca_2024, BHEX_Lupsasca_2024, BHEX_Galison_2024, BHEX_Kawashima_2024}.
Furthermore, BHEX will enable demographic studies of event-horizon-scale properties for at least a dozen additional supermassive black holes (SMBHs) currently known\cite{Akiyama_PASP_2026}, yielding crucial insights into the processes that drive their formation and growth.
Additionally, BHEX will connect these SMBHs to their relativistic jets, shedding light on the mechanisms that power the universe's most luminous and efficient engines.

To address these scientific questions, BHEX will explore the universe at the highest angular resolution ever achieved in astronomy.
It will offer transformative scientific capabilities that are not accessible with terrestrial telescopes alone by extending existing ground-based millimeter/submillimeter VLBI networks---such as the Event Horizon Telescope (EHT) and the Global Millimeter VLBI Array (GMVA)---into space.
BHEX is configured as a dual-band instrument that can simultaneously capture signals in two frequency bands\cite{BHEX_Johnson_2024, BHEX_Marrone_2024}, enabling the use of the frequency phase transfer (FPT) technique\cite{Asaki_1996, Dodson_2009, Rioja_2011}.
The receiver operating ranges are the 100\,GHz band (80--106\,GHz) and the 300\,GHz band (240--320\,GHz)\cite{BHEX_Marrone_2024, BHEX_Tong_2024}, which overlap with the bands used by the GMVA and EHT, respectively.
The foundational BHEX mission concept was detailed in a series of papers presented at the previous edition of this conference\cite{BHEX_Akiyama_2024, BHEX_Johnson_2024, BHEX_Marrone_2024, BHEX_Peretz_2024, BHEX_Lupsasca_2024, BHEX_Galison_2024, BHEX_Issaoun_2024, BHEX_Kawashima_2024, BHEX_Tomio_2024, BHEX_Wang_2024, BHEX_Sridharan_2024, BHEX_Rana_2024, BHEX_Tong_2024, BHEX_Srinivasan_2024}.

The Japanese astronomical community has been a major contributor to the development of the BHEX mission concept---from scientific goals to instrument design---through the BHEX Japan Consortium, which was formed in 2023\cite{BHEX_Akiyama_2024}.
BHEX represents a natural extension of ongoing VLBI research in Japan and aligns closely with the community's strategic priorities for the 2030s\cite{Akahori_2021}.
The mission provides a clear opportunity to address many key scientific objectives in active galactic nucleus (AGN) and SMBH research identified by the Japanese community.
The BHEX Japan Consortium aims to advance the mission as a collaborative space program between the United States and Japan.
As a continuation of our 2024 contribution\cite{BHEX_Akiyama_2024}, this paper briefly outlines subsequent developments and major accomplishments in Japan. It is part of a series of contributions reporting on the mission's development\cite{BHEX_Johnson_2026, BHEX_Houston_2026, BHEX_Gray_2026}.
We briefly introduce the BHEX Japan Consortium in \autoref{sec:bhex_japan}, followed by a summary of major progress since 2024 in \autoref{sec:bhex_japan_updates}.
Finally, we summarize our future outlook in \autoref{sec:summary}.

\section{The BHEX Japan Consortium}
\label{sec:bhex_japan}
As described in Akiyama et al.\cite{BHEX_Akiyama_2024}, the BHEX Japan Consortium was formed in September 2023 to investigate the mission's scientific goals and to define potential Japanese contributions to the BHEX instrument and its operations.
As of June 2026, the Consortium includes more than 60 scientists from over 25 institutions, with expertise spanning observational radio astronomy (single-dish and interferometric, including but not limited to VLBI), astronomical instrumentation and technology development, theoretical physics/astrophysics, and optical laser communications.
The Consortium and its subgroups meet regularly to refine Japanese contributions while maintaining close coordination with the broader U.S.-led BHEX collaboration.

The BHEX Japan science team is organized into three science working groups, each targeting distinct mission-enabled opportunities\cite{BHEX_Akiyama_2024, BHEX_Kawashima_2024}. The General Relativity, Accretion, and Jet Launching Working Group focuses on event-horizon-scale physics, using BHEX's ultra-high angular resolution to resolve the black hole photon ring and to measure properties of the black hole spacetime and magnetosphere that are otherwise inaccessible. The AGN Working Group expands the horizon-scale target sample and investigates the universality of jet launching and particle-acceleration mechanisms across a diverse population of active galactic nuclei. The Single Dish Working Group explores a potential standalone observing mode that uses expected mission downtime to conduct high-sensitivity searches for interstellar molecular oxygen and to perform line surveys of the molecular universe in the otherwise atmosphere-blocked 50--70\,GHz band.

To support the mission's baseline fringe-sensitivity requirement of $\sim 1\,\mathrm{mJy}$\cite{BHEX_Johnson_2024, BHEX_Marrone_2024, BHEX_Peretz_2024}, the BHEX Japan Consortium has identified primary technical and instrumental contributions\cite{BHEX_Akiyama_2024}, including the three contributions outlined below. These contributions are strategically divided between a sensitive broadband receiver system on the BHEX satellite\cite{BHEX_Tong_2024} and terrestrial infrastructure to support ground--space VLBI observations.
\begin{itemize}
    \item \textbf{300\,GHz SIS Mixer:} Supplying high-critical-current-density Superconductor--Insulator--Superconductor (SIS) mixer devices for the onboard dual-polarization 240--320\,GHz receiver system, developed by the Advanced Technology Center (ATC) of the National Astronomical Observatory of Japan (NAOJ). This builds on heritage from the Superconducting Submillimeter-Wave Limb-Emission Sounder (SMILES) mission\cite{Masuko_1997} led by the Japan Aerospace Exploration Agency (JAXA), as well as decades of SIS-mixer development at NAOJ ATC for ground facilities such as the Atacama Large Millimeter/submillimeter Array (ALMA) and its planned wideband upgrade.
    \item \textbf{4.5\,K Cryocooler:} Providing a space-qualified, multistage, fully closed-cycle cryocooling system\cite{BHEX_Rana_2024, Rana_2026} manufactured by Sumitomo Heavy Industries (SHI) to maintain the 4.5\,K bath temperature required for optimal SIS mixer performance. Japan has developed world-leading expertise in space cryocooler technology at these temperatures; to date, only three $\sim$4\,K fully closed-cycle cryocooler systems have successfully operated in space, all manufactured by SHI for JAXA scientific missions. SHI has developed space cryocooler technologies since 1987, and this work leverages heritage from JAXA missions including SMILES\cite{Inatani_2005, Otsuka_2010, Ochiai_2010, Narasaki_2012}, Hitomi\cite{Yoshida_2018, Fujimoto_2018}, and the X-ray Imaging and Spectroscopy Mission (XRISM)\cite{Ezoe_2020, Imamura_2023}.
    \item \textbf{Ground Millimeter/Submillimeter Observatories:} Enhancing the ground-based VLBI network and anchor-station coverage at the 100\,GHz band by integrating facilities operated by NAOJ, including the VLBI Exploration for Radio Astrometry (VERA; being extended to 86\,GHz) of the Mizusawa VLBI Observatory, and the Nobeyama 45-m Telescope at the Nobeyama Radio Observatory, which are both a part of the greater East Asian VLBI Network (EAVN)\cite{Akiyama_2022}.
\end{itemize}

\section{Progress in the Mission Development in Japan}
\label{sec:bhex_japan_updates}

\subsection{BHEX Working Group}
A major milestone for BHEX Japan was the establishment of the Black Hole Explorer Working Group (BHEX WG) in June 2025 under the Advisory Committee for Space Science at the Institute of Space and Astronautical Science (ISAS), JAXA.
JAXA has supported BHEX mission development in Japan since then through the BHEX WG.
In Japan, a working group under this committee is a formal pathway for the community to conduct Pre-Phase~A mission studies for a future space science mission concept.
The establishment of the BHEX WG was proposed by the BHEX Japan Consortium in late 2024 after a year of initial mission concept studies.

The research and development activities of the BHEX WG have focused primarily on two key onboard instruments: a 4.5\,K closed-cycle mechanical cryocooler and an ultra-wideband 300\,GHz SIS mixer. Here we outline major progress in these two areas:\vspace{0.5em}\\
\textbf{Cryocooler:} The establishment of the BHEX WG enables the BHEX Japan community to engage the space cryocooler group at ISAS/JAXA, leveraging decades of expertise from past JAXA missions and the broader BHEX community. The BHEX WG completed concept design studies of the BHEX cryocooler with SHI, the cryocooler vendor for the JAXA Hitomi, XRISM, and SMILES missions. The studies produced a cryocooler concept design that meets the BHEX requirements \cite{BHEX_Rana_2024, Rana_2026} and uses only commercial off-the-shelf components with on-orbit flight heritage on XRISM and past JAXA missions, exceeding the two-year lifetime anticipated for a 2026 NASA Astrophysics Small Explorers mission. The BHEX cryocooler concept design was integrated into the overall BHEX receiver system design\cite{BHEX_Tong_2024}.\vspace{0.5em}\\
\textbf{SIS Mixer:}
With BHEX WG financial support, an initial prototype 300\,GHz SIS mixer was fabricated at NAOJ ATC for the BHEX mission. The mixer employs a three-junction series SIS array, a waveguide probe with a microstrip impedance transformer, and an on-chip IF tuning circuit.
As a first implementation of the baseline RF design, the mixer chip was fabricated using an established aluminum-oxide-barrier (Nb/Al--AlOx/Al/Nb) SIS junction process, following the fabrication process of ALMA Band 8 receivers \cite{Tamura_2015}.
Evaluated at a mixer-block temperature of 4\,K, the fabricated chip achieved a double-sideband receiver noise temperature of 26--41\,K at LO frequencies of 240--307\,GHz planned for BHEX \cite{BHEX_Tong_2024}, close to the two-photon sensitivity target adopted for the BHEX 300\,GHz receiver \cite{BHEX_Johnson_2024,BHEX_Marrone_2024}. These results, reported in Murayama et al.~\cite{Murayama_2026}, demonstrate low-noise operation and provide feedback to improve RF/IF bandwidth and mixer stability in subsequent designs.

\subsection{Ground Stations}
The persistent impact of EHT results over the last several years, the advent of stations operating at 86\,GHz in East Asia\cite{Akiyama_2022}, and growing community motivation to expand existing facilities (including BHEX) have driven recent upgrades to millimeter facilities in Japan, including VERA and the Nobeyama 45\,m Telescope. Here we summarize recent progress at these two ground stations.

\subsubsection{Nobeyama 45\,m Telescope}
Operating up to $\sim 100\,$GHz, the Nobeyama 45\,m Telescope remains among the largest millimeter-wave single-dish telescopes. It has contributed to global 86\,GHz VLBI since 1988 \cite{Baath_1991,Baath_1992} and participates in the Japanese VLBI Network (JVN) and EAVN \cite{Akiyama_2022}.
Through the HINOTORI receiver upgrade (Hybrid Integration Project in Nobeyama, Triple-band Oriented), the telescope now supports simultaneous 22/43/86\,GHz observations in both single-dish and VLBI modes \cite{Okada_2020, Tsutsumi_2023, Imai_2023}.
With this capability, Nobeyama is currently the most sensitive 86\,GHz VLBI station in East Asia and is well positioned to serve as an anchor station for BHEX at $\sim 100\,$GHz.

The 45\,m Telescope is also expected to offer attractive sensitivity at 230\,GHz. Previous holographic measurements indicate an RMS surface error of $\sim 80\,\mu\mathrm{m}$ (and $\sim 54\,\mu\mathrm{m}$ for the inner 40\,m).
This corresponds to an anticipated aperture efficiency of $\gtrsim 20\,\%$ at 230\,GHz, equivalent to the collecting area of a 26\,m dish with $\gtrsim 60\,\%$ efficiency (cf. the 15\,m James Clerk Maxwell Telescope with $\sim 60\,\%$ aperture efficiency).
As a millimeter-wave radio astronomy site, Nobeyama typically provides a winter atmospheric opacity of $\sim 0.2$--$0.4$ \cite{Nishimura_2020}.

Higher-frequency capabilities beyond 150\,GHz have also been explored over the telescope's history. From the late 1990s to the early 2000s, a series of 230\,GHz experiments was carried out to expand the Nobeyama Rainbow Interferometer \cite{Shibatsuka_2001}. This mode incorporated the 45\,m Telescope into the Nobeyama Millimeter Array (NMA) with its six 10-m antennas \cite{Hagiwara_1998, Momose_1999a, Momose_1999b} and was also called the ``Rainbow'' mode of the NMA (see, e.g., Refs.~\cite{Hagiwara_1999,Imanishi_2004,Imanishi_2006,Imanishi_2007,Kohno_2008,Okuda_2005,Okuda_2013,Sofue_2001,Yamada_2007,Furuya_2002,Iono_2006}).

The strong potential of the 45\,m Telescope as an East Asian anchor station for 230\,GHz VLBI has motivated a five-year receiver-upgrade program to enable 230\,GHz VLBI, starting in 2026 with financial support from the Japan Society for the Promotion of Science. The program scope includes a multi-frequency upgrade of the telescope's TZ receiver, a two-beam waveguide-type, dual-polarization, sideband-separating SIS receiver system in the 100\,GHz band \cite{Nakajima_2013}. The TZ receiver will be upgraded to support simultaneous dual-frequency reception in the 100\,GHz and 230\,GHz bands for frequency phase transfer (FPT) by replacing the second beam with a dedicated waveguide, horn, and SIS mixer for 230\,GHz.

A prototype 230\,GHz engineering beam system was installed in the TZ receiver and evaluated at the 45\,m Telescope in 2026. Over the next few years, the project will further refine the 230\,GHz beam system design and implement a dedicated VLBI backend system to meet the requirements for dual-band observations for the EHT and BHEX, targeting first participation in EHT observations around 2029, a few years before the planned launch of BHEX in 2031.

\subsubsection{VERA}
VERA is a Japanese VLBI array comprising four 20-m telescopes (Iriki, Ishigakijima, Mizusawa, and Ogasawara) \cite{Honma_2000}. It was primarily designed for precise VLBI astrometry of Galactic maser sources to study the structure of the Milky Way. Although VERA has operated primarily at 6.7\,GHz, 22\,GHz, and 43\,GHz, each antenna was designed with the potential to extend operations to 86\,GHz\cite{Honma_2000}.

86\,GHz VLBI experiments were recently conducted in December 2024 on the 2,300\,km baseline between the northern Mizusawa and southwestern Ishigaki stations using engineering room-temperature receivers. The experiments achieved successful fringe detections, confirming aperture efficiencies of $\sim 29$\,\% and adequate atmospheric transmission. A new dual-polarization cryogenic 86\,GHz receiver has now been developed for VERA to significantly expand the 86\,GHz network of the EAVN \cite{Kameyama_2024}. It is planned for installation at the Mizusawa station in late 2026, followed by the other VERA stations over the next few years.

\subsection{Community Training and Engagement}
\begin{figure}
    \centering
    \includegraphics[width=0.9\textwidth]{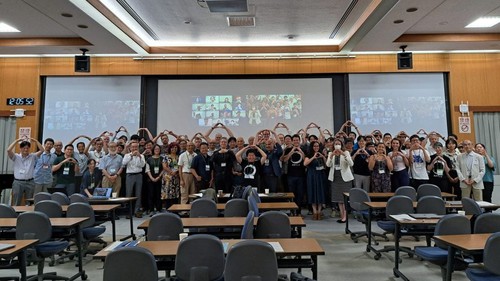}
    \caption{A group photo from ``The Black Hole Explorer Japan Workshop 2024,'' held at the National Astronomical Observatory of Japan in June 2024.}
    \label{fig:bhexjapanworkshop}
\end{figure}

The overall BHEX mission concept and mission development in Japan have been disseminated regularly to domestic, regional, and global communities through national and international conferences, including the 2024 and 2026 editions of this conference series\cite{BHEX_Akiyama_2024, BHEX_Kawashima_2024}.
We also hosted the first international conference focused on BHEX, ``The Black Hole Explorer Japan Workshop 2024''\footnote{\url{https://sites.mit.edu/bhex-japan-workshop-2024/}, accessed on June 30, 2026} at NAOJ in June 2024 (\autoref{fig:bhexjapanworkshop}). The conference brought together $\sim$160 scientists from around the world, with approximately half from East Asian communities, and highlighted the broad dissemination of the BHEX mission within Japan and the region.
BHEX Japan also co-hosted the online conference ``BHEX Fall 2025 Science Meeting''\footnote{\url{https://www.blackholeexplorer.org/events/bhex-fall-2025-science-meeting}, accessed on June 30, 2026} with the broader BHEX community.

\begin{figure}
    \centering
    \includegraphics[height=0.9\textheight]{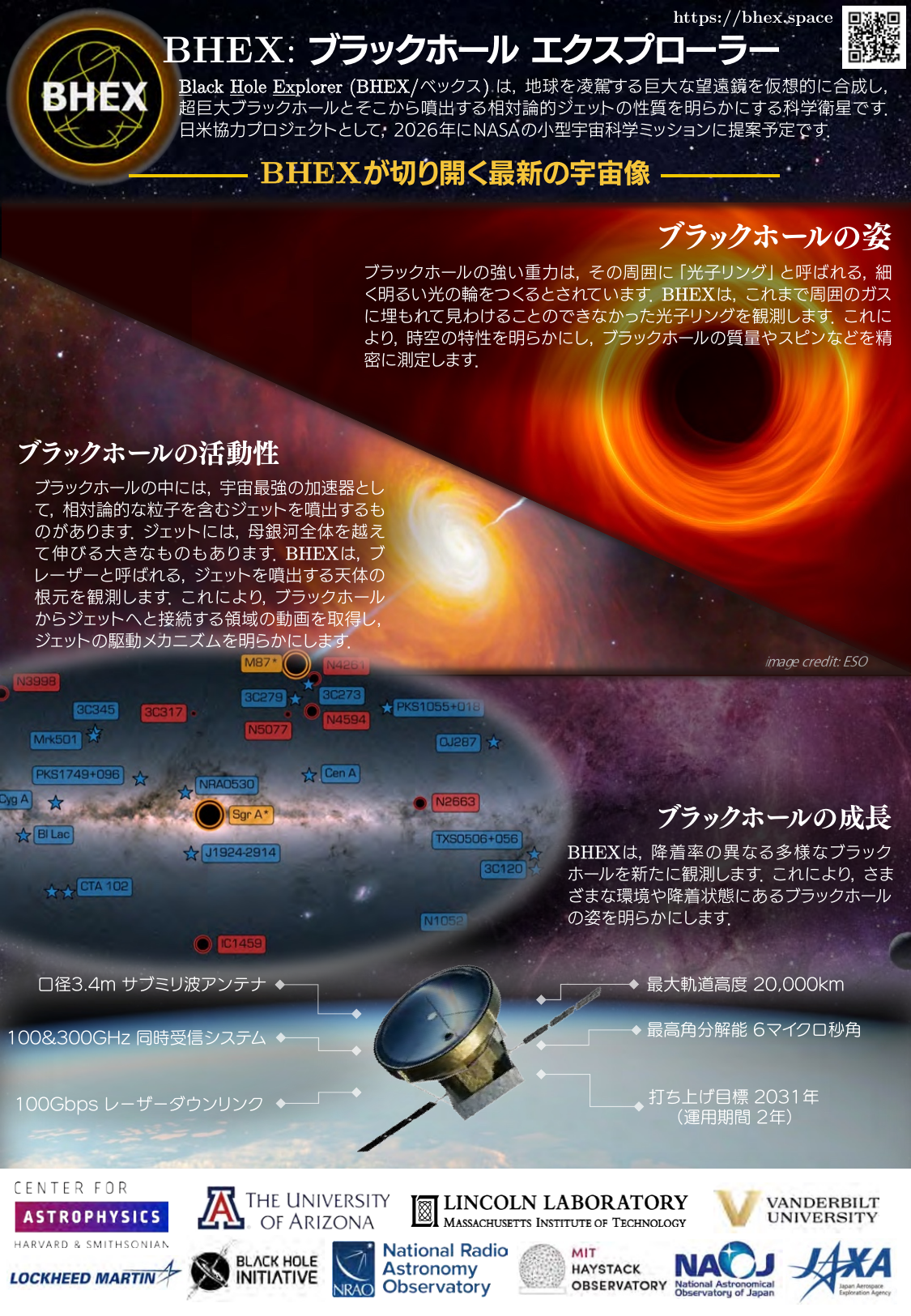}
    \caption{The latest fact sheet in Japanese (accessed on June 30, 2026). The fact sheet is available in multiple languages on the BHEX website.}
    \label{fig:factsheet}
\end{figure}
The BHEX Japan community has contributed to developing the science goals and operations concept. Early science concepts were presented in Akiyama et al.~\cite{BHEX_Akiyama_2024} and Kawashima et al.~\cite{BHEX_Kawashima_2024}. The community also plans to present additional science cases for BHEX in the ongoing special issue, ``The Black Hole Explorer,'' in the Publications of the Astronomical Society of the Pacific\footnote{\url{https://iopscience.iop.org/collections/pasp-260609-01}, accessed on June 30, 2026}, including detailed observational simulations of BHEX observations of new event-horizon-scale SMBH targets in nearby low-luminosity AGNs \cite{Akiyama_PASP_2026}.
For the broader community, BHEX Japan released a Japanese version of the mission fact sheet on the BHEX website\footnote{\url{https://www.blackholeexplorer.org/fact-sheet}, accessed on June 30, 2026} (\autoref{fig:factsheet}).

Even at this early stage of the mission, BHEX Japan has provided opportunities to train the next generation of scientists, who currently make up about one-third of the BHEX Japan Consortium \cite{BHEX_Akiyama_2024}. Since 2024, BHEX Japan has exchanged multiple early-career scientists with U.S. partner institutions through bilateral U.S.--Japan mission-development programs, resulting in one bachelor's thesis and one master's thesis within BHEX Japan during the past year alone. We also anticipate completion of several additional theses focused on BHEX mission studies over the next few years, as international collaboration with BHEX scientists worldwide accelerates.

\section{Future Outlook}
\label{sec:summary}
We have outlined the current status of BHEX mission development in Japan.
The BHEX Japan Consortium has actively participated in the ongoing development of the mission concept.
As a result of three years of activity in Japan, Pre-Phase~A studies are now supported by JAXA through the BHEX WG under the Advisory Committee for Space Science of ISAS/JAXA.
This support has accelerated the BHEX cryocooler concept design and the development of the broadband 300\,GHz SIS mixer for BHEX.
Community motivation for EHT and BHEX science has also driven upgrades to ground-based Japanese facilities, including VERA and the Nobeyama 45\,m Telescope, to support 86\,GHz VLBI and simultaneous 100+230\,GHz VLBI, respectively.
In close collaboration with the broader BHEX community, the Consortium is contributing to a proposal for the 2026 NASA Astrophysics Small Explorers (SMEX) solicitation, due in fall 2026.

\acknowledgments %
The mission concept studies for BHEX within the BHEX Japan Consortium have been financially supported by the  programs and organizations, including the followings:
\begin{itemize}
\item Strategic Development Funding for the BHEX Working Group under the Advisory Committee for Space Science, ISAS/JAXA
\item ULVAC-Hayashi Seed Fund from the MIT-Japan Program at MIT International Science and Technology Initiatives (MISTI)
\item MEXT/JSPS Grants-in-Aid for Scientific Research (KAKENHI) Grant Numbers: JP15H00784 (K.N.), JP15H03644 (Y.H.), JP19KK0081 (M.H), JP17K05398 (T.H.), JP18KK0090 (K.H.), JP18H03721 (K.N.), JP19H01943 (K.H., Y.H., K.N.), JP21H04488 (K.O.), JP21K03628 (K.N.), JP22H00157 (K.H., S.K., K.N.), JP22H04955 (Y.U.), JP23H00117 (T.K.), JP23H00118 (K.N.), JP23K03448 (T.K.), JP23K03453 (S.K.), JP24684011 (T.H.), JP25K07357 (Y.H.), JP26K17203 (Y.M.), JP26K21725 (K.H.)
\item MEXT, ``Program for Promoting Research on the Supercomputer Fugaku'' (Structure and Evolution of the Universe Unraveled by Fusion of Simulation and AI; Grant Number JPMXP1020240219 (K.O., T.K.); Black hole accretion disks and quasi-periodic oscillations revealed by general relativistic hydrodynamics simulations and general relativistic radiation transfer calculations; Grant Number JPMXP1020240054 (K.O., T.K.))
\item The Joint Institute for Computational Fundamental Science (JICFuS; K.O.)
\item The Multidisciplinary Cooperative Research Program in CCS, University of Tsukuba (K.O.)
\item JST Moonshot R\&D Grant Number JPMJMS2067 (Y.U.)
\end{itemize}
In addition to the above programs, the BHEX mission concept studies have been supported by funding from organizations including:
the Smithsonian Astrophysical Observatory,
the Internal Research and Development (IRAD) program at NASA Goddard Space Flight Center,
the University of Arizona,
the Black Hole Initiative at Harvard University, 
the National Science Foundation (AST-2307887),  
and the Gordon and Betty Moore Foundation (Grant \#13526).
It was also made possible through the support of a grant from the John Templeton Foundation (Grant \#63445).  The opinions expressed in this publication are those of the author(s) and do not necessarily reflect the views of these Foundations. 
BHEX is funded in part by generous support from Mr. Michael Tuteur and Amy Tuteur, MD. 
BHEX is supported by initial funding from Fred Ehrsam.

\end{document}